\documentclass[twocolumn,prd,showpacs,nofootinbib,epsfig]{revtex4}
\usepackage{epsfig}
\def\beq{\begin{equation}}
\def\eeq{\end{equation}}
\def\bea{\begin{eqnarray}}
\def\eea{\end{eqnarray}}
\def\bq{\begin{quote}}
\def\eq{\end{quote}}

\def\nnb{\nonumber}
\def\ga{\left(}
\def\dr{\right)}

\def\nnb{\nonumber}
\def\la{\langle}
\def\ra{\rangle}
\def\nin{\noindent}
\def\ba{\vspace*{-0.2cm}\begin{array}}
\def\ea{\end{array}\vspace*{-0.2cm}}

\def\b{$\bullet~$}
\def\als{\alpha_s}

\def\gg2{ \la\alpha_s G^2 \ra}
\def\gg3{g^3f_{abc}\la G^aG^bG^c \ra}
\def\ggg4{\la\als^2G^4\ra}

\def\beq{\begin{equation}}
\def\enq{\end{equation}}
\def\beqa{\begin{eqnarray}}
\def\enqa{\end{eqnarray}}
\def\nnb{\nonumber}

\def\GeV{\nobreak\,\mbox{GeV}}

\def\qq{\lag\bar{q}q\rag}

\def\mix{\lag\bar{q}g\si.Gq\rag}

\def\Gd{\lag g^2G^2\rag}
\def\G3{\lag g^3G^3\rag}

\def\rh{\rho}
\def\si{\sigma}

\def\al{\alpha}
\def\be{\beta}
\def\alma{\alpha_{max}}
\def\almi{\alpha_{min}}
\def\bemi{\beta_{min}}
\def\lb{\label}
\def\nn{\nonumber}
\def\bd{\boldmath}
\def\kab{\left[(\al+\be)m_c^2-\al\be s\right]}
\newcommand{\rag}{\rangle}
\newcommand{\lag}{\langle}

\begin{document}

\title{On the nature of the X(3872) from QCD}

 \author{S. Narison}
\email{snarison@yahoo.fr}
\affiliation{Laboratoire
de Physique Th\'eorique et Astroparticules, CNRS-IN2P3 and Universit\'e
de Montpellier II, 
\\
Case 070, Place Eug\`ene
Bataillon, 34095 - Montpellier Cedex 05, France.}
 \author{F. S.  Navarra}
\email{navarra@if.usp.br}
\affiliation{Instituto de F\'{\i}sica, Universidade de S\~{a}o Paulo, 
C.P. 66318, 05389-970 S\~{a}o Paulo, SP, Brazil}
\author{M. Nielsen}
\email{mnielsen@if.usp.br}
\affiliation{Instituto de F\'{\i}sica, Universidade de S\~{a}o Paulo, 
C.P. 66318, 05389-970 S\~{a}o Paulo, SP, Brazil}

\begin{abstract}
\nin
We have studied some possible four-quark and molecule configurations of the 
$X(3872)$ 
using double ratios of sum rules, which are more accurate than the usual simple 
ratios 
often used in the literature to obtain  hadron masses. We found that the different  
structures ($\bar 3-3$ and $\bar 6-6$  tetraquarks and $D-D^{(*)}$ molecule) lead to 
the same prediction for the mass (within the accuracy of the method), indicating that 
the alone prediction of the  $X$  mass may not be sufficient to reveal its  nature. 
In doing these analyses, we also find that  (within our approximation) the use of the 
$\overline{MS}$ running ${\overline m}_c(m_c^2)$, rather than the on-shell mass, 
is more appropriate to obtain  the $J/\psi$ and $X$ meson masses.
Using vertex sum rules to roughly estimate the $X(3872)$ hadronic and radiative 
widths, we found that the ones of a  $\lambda-J/\psi$-like molecule current can be compatible within the errors with the data.
\end{abstract}
\maketitle

\section{Introduction}
\vspace*{-0.2cm}
\nin
The nature of the narrow ($\leq$ 2.3 MeV width) $X(3872)$ decaying to $J/\psi\pi^
+\pi^-$ \cite{PDG} discovered by BELLE in $B$-decays \cite{BELLE}
and confirmed by BABAR \cite{BABAR}, CDF \cite{CDF} and D0 \cite{D0}
in hadronic productions remains puzzling. Different scenarios (four-quark state, 
molecule,
large mixing with conventional $\bar cc$ states) have been evoked in the 
literature \cite{Swanson,rev}.
In this work we use QCD spectral sum rules (QSSR) (the Borel/Laplace Sum
Rules \cite{svz,rry,SNB}) in order to test the previous four-quark and molecule 
scenarios.
\\
In a previous calculation \cite{mnnr} some of us and our collaborators have 
considered the
$X(3872)$ as being a tetraquark state where the diquark-antidiquark pairs are in 
the $\bar{3}-3$ color configuration.  
A priori, the diquark-antidiquark pairs could also be in a $\bar 6-{6}$ color 
configuration. This  system is expected 
to be too weakly  bound  by a two-body potential but it could be bound by a 
four-body potential, such as the one of the 
Steiner model \cite{RICHARD}. In this work we shall, for the first time,  
investigate this configuration using QSSR. \\
Using QSSR, we shall, also for the first time, analyze  the mass and hadronic 
width of a $\lambda-J/\psi$-like molecule  
\footnote{An analogous configuration has been studied  within QSSR for light 
four-quark states in \cite{SN4}.}, 
which we shall compare with the ones of the four-quark states. 
\section{The interpolating $X$-currents}
\nin
 In order to study the two-point functions of the $X(3872)$ meson
 assumed to be an $1^{++}$ axial
vector meson,  
The interpolating current which describes the $X(3872)$ as a diquark-antidiquark  
system 
in the $\bar{3}-3$ color configuration with total $J^{PC}=1^{++}$ is \cite{mnnr}:
\bea
j^\mu_{3}&=&{i\epsilon_{abc}\epsilon_{dec}\over\sqrt{2}}[(q_a^TC\gamma_5c_b)
(\bar{q}_d\gamma^\mu C\bar{c}_e^T)\nnb\\
&&+(q_a^TC\gamma^\mu c_b)
(\bar{q}_d\gamma_5C\bar{c}_e^T)]\;,
\label{field3}
\eea
while for a
diquark-antidiquark in the color sextet ($\bar 6-6$) configuration, the 
interpolating current is:
\bea
j^\mu_{6}&=&{i\over\sqrt{2}}[(q_a^TC\gamma_5\lambda^S_{ab}c_b)
(\bar{q}_d\gamma^\mu C\lambda^S_{de}\bar{c}_e^T)\nnb\\
&&+(q_a^TC\gamma^\mu
\lambda^S_{ab} c_b)(\bar{q}_d\gamma_5C\lambda^S_{de}\bar{c}_e^T)]\;,
\label{field}
\eea
where $a,~b,~c,~...$ are color indices, $C$ is the charge conjugation
matrix, $q$ denotes a $u$ or $d$ quark and $\lambda^S$ stands for the 
six symmetric Gell-Mann matrices:
$\lambda^S=\left(\lambda_0,~\lambda_1,~\lambda_3,~\lambda_4,~\lambda_6,
~\lambda_8\right)$. \\
These tetraquark currents can be compared with the one describing the
$X$ state as a $D^*-D$ molecule:
\bea
j_{mol}^{\mu}(x) & = & \ga{g\over\Lambda}\dr^2_{\rm eff}{1 \over \sqrt{2}}
\Big{[}
\left(\bar{q}_a(x) \gamma_{5} c_a(x)
\bar{c}_b(x) \gamma^{\mu}  q_b(x)\right)\nnb\\
&&- \left(\bar{q}_a(x) \gamma^{\mu} 
c_a(x)\bar{c}_b(x) \gamma_{5}  q_b(x)\right)
\Big{]}.
\lb{curr4}
\eea
and as a $\lambda-J/\psi$-like molecule current:
\bea
j^{\mu}_{\lambda}  =  \ga{g'\over\Lambda}\dr^2_{\rm eff}
\left(\bar{c}\lambda^a\gamma^{\mu}c\dr
\ga\bar{q}\lambda_a  \gamma_5 q\right)
\lb{eq:curr5}
\eea
where $\lambda_a$ is the colour matrix. \\
In the molecule assignement  it is assumed that there is an effective local current 
and  the  meson pairs are weakly 
bound by a van der Vaals force in a Fermi-like theory with a strength 
$(g/\Lambda)^2_{\rm eff}$ which has nothing to 
do with the quarks and gluons inside each meson. 
\section{The  two-point correlator and form of the sum rules}
\nin
%
The two-point correlation function associated to the axial-vector current is 
defined as:
\bea
\Pi^{\mu\nu}_i(q)&=&i\int d^4x ~e^{iq.x}\lag 0
|T[j^\mu_{i}(x){j^\nu_{i}}^\dagger(0)]
|0\rag\nnb\\
&=&-\Pi_{1i}(q^2)(g^{\mu\nu}-{q^\mu q^\nu\over q^2})+\Pi_{0i}(q^2){q^\mu
q^\nu\over q^2},
\lb{2po}
\eea
where $i=3,~6,~mol,~\lambda$ according to the currents in Eqs.~(\ref{field3}),
(\ref{field}), (\ref{curr4}) and (\ref{eq:curr5}). The two functions,
$\Pi_1$ and $\Pi_0$, appearing in Eq.~(\ref{2po}) are independent and
have respectively the quantum numbers of the spin 1 and 0 mesons.
\\
Due to its analyticity, the correlation function, $\Pi_{1i}$, obeys a
dispersion relation:
\beq
\Pi_{1i}(q^2)=\int_{4m_c^2}^\infty ds {\rho_i(s)\over s-q^2}+...\;,
\lb{ope}
\enq
where $\pi \rho_i(s)\equiv\mbox{Im}[\Pi_{1i}(s)]$ is the spectral function.  
After
making an inverse-Laplace (or Borel) transform on both sides, the sum rule
and its ratio read:
\bea {\cal F}_i(\tau)&=&\int_{4m_c^2}^{\infty}ds~
e^{-s\tau}~\rho_i(s)\; \nnb\\
{\cal R}_i(\tau)&=& -{d\over d\tau}{\log {\cal F}_i(\tau)}~,
\lb{sr1} 
\eea
where $\tau\equiv 1/M^2$ is the sum rule variable with $M$ being the 
inverse-Laplace (or Borel) mass. 
In the following, we shall work with the double ratio of sum rules (DRSR):
\beq
r_{ij}= \sqrt{{\cal R}_i\over {\cal R}_j}~~: i=3,6,...
\eeq
to obtain the $X$-meson mass. 
Defining the coupling of the current with the state through:
\beq\label{eq: decay}
\lag 0 |
j^\mu_{i}|X\rag =\sqrt{2}f_XM_i^4\epsilon^\mu~, 
\enq
and using the minimal duality ansatz: ``one resonance" $\oplus$ ``QCD continuum", 
where the QCD
continuum comes from the discontinuity of the QCD diagrams from a continuum 
threshold $t_c$, the phenomenological side of Eq.~(\ref{2po}) can be written as:
\beq
\Pi_{\mu\nu}^{phen}(q^2)={2f_X^2M_i^8\over
M_i^2-q^2}\left(-g_{\mu\nu}+ {q_\mu q_\nu\over M_i^2}\right)
+\cdots\;, \lb{phe} \enq
where the Lorentz structure projects out the $1^{++}$ state.
The dots
denote higher axial-vector resonance contributions that will be
parametrized, as usual, by the QCD continuum. 
Transferring the continuum contribution to the QCD side, the sum rules
can be written in a finite energy form as: 
\bea {\cal F}_i(\tau)&\equiv& 2f_X^2M_i^8e^{-M_i^2\tau}=\int_{4m_c^2}^{t_c}ds~
e^{-s\tau}~\rho_i(s)\; \nnb\\
{\cal R}_i(\tau)&\equiv& -{d\over d\tau}{\log {\cal F}_i(\tau)}\simeq M_i^2~,\nnb\\
r_{ij}&\equiv& \sqrt{{\cal R}_i\over {\cal R}_j}\simeq {M_i\over M_j}~~~~~:~~~ 
i=3,6,...
\lb{sr} \eea

%
\section{The  QCD expressions of the two-point correlators}
\nin
The QCD expressions of the spectral densities of the two-point correlator 
associated to the 
currents in Eqs. (\ref{field3}) and (\ref{curr4}) have been obtained
respectively in \cite{mnnr} and \cite{x24} and will not be reported here.
The expression associated to the current in Eqs. (\ref{field}) and (\ref{eq:curr5}) 
are new. Up to dimension-six condensates, we can write:
\bea
\rho_i(s)&=&\rho_i^{pert}(s)+\rh_i^{m_q}(s)+\rh_i^{\qq}(s)+\nnb\\
&&\rh_i^{\lag G^2\rag}
(s)+\rh_i^{mix}(s)+\rh_i^{\qq^2}(s)\;.
\lb{rhoeq}
\eea
The renormalization improved perturbative expression of the sum rule is given by:
\beq
{\cal F}_i(\tau)|_{pet}=(\alpha_s(\tau))^{-{\gamma_i\over \beta_1}}A_i\Big{[}1+
K_i{\alpha_s\over \pi}+\cdots\Big{]}~,
\eeq
where $\gamma_i$ is the anomalous dimension of the corresponding correlator, 
$-\beta_1=(1/2)(11
-2n/3)$ is the first coefficient of the $\beta$-function for SU(n) flavours,
$A_i$ is the known LO expression and  $K_i$ is the radiative correction. By 
inspection we observe 
that in the ratio of moments ${\cal R}$ defined in 
Eq. (\ref{sr}), the $\alpha_s$ corrections disappear and  only  the radiative 
corrections induced 
by the anomalous dimensions of the currents survive.   
In the double ratios of sum rules (DRSR) which we shall use in this paper, this 
induced
radiative correction will also disappear to ${\cal O}(\alpha_s)$ as the different 
currents studied 
(which have all the same Lorentz structure) have the same anomalous dimensions. 
Therefore, we expect that, 
although we work in leading order of the QCD expressions, our results for the 
ratios of 
masses are accurate up to order $\alpha_s$ for the perturbative contributions.  \\
%
\subsection{ \bf \bd$6-6$ four-quark current}
\nin
For the $6-6$ current in Eq.~(\ref{field}), we get to lowest order in $\alpha_s$:
%
\beqa
\label{eq:pert}
\rho_6^{pert}(s)&=&{1\over 2^{9} \pi^6}\int\limits_{\almi}^{\alma}
{d\al\over\alpha^3}
\int\limits_{\bemi}^{1-\al}{d\be\over\be^3}(1-\al-\be)\nnb\\
&&\times(1+\al+\be)\left[(\al+\be)
m_c^2-\al\be s\right]^4,
\nn\\
\rho_6^{m_q}(s)&=&-{m_q \over 2^2 \pi^4} \int\limits_{\almi}^{\alma}
{d\al\over\al} 
\bigg\{ -{\qq\over 2^2}{[m_{c}^2-\al(1-\al)s]^2 \over (1-\al)}\nnb\\
&&+ \int\limits_{\bemi}^{1-\al}{d\be\over\be}\kab \bigg[ 
- m_{c}^{2} \qq\nn\\
&&+{\qq\over 2^2} \kab
\nn\nnb\\
&&+{m_c\over 2^5 \pi^2 \al \be^2}(3+\al+\be)(1-\al-\be)\nnb\\
&&\times\kab^2 
\bigg] \bigg\},\nn\\
\rho_6^{\qq}(s)&=&-{m_c\qq\over 2^{4}\pi^4}\int\limits_{\almi}^{\alma}
{d\al\over\al^2}
\int\limits_{\bemi}^{1-\al}{d\be\over\be}(1+\al+\be)\nnb\\
&&\times\left[(\al+\be)m_c^2-
\al\be s\right]^2,\nn\\
\rho_6^{\la G^2\ra}(s)&=&{\Gd\over2^{8}3\pi^6}\int\limits_{\almi}^{\alma} d\al\!\!
\int\limits_{\bemi}^{1-\al}{d\be\over\be^2}\left[(\al+\be)m_c^2-\al\be
s\right]\nnb\\
&&\times\Bigg{[}{m_c^2(1-(\al+\be)^2)\over\be}+
{(1-2\al-2\be)\over4\al}\nnb\\
&&\times\Big{[}(\al+\be)m_c^2-\al\be s\Big{]}
\Bigg]. \nnb
\enqa
\beqa
\rho_6^{mix}(s)&=&{m_c\mix\over 2^{5}\pi^4}\int\limits_{\almi}^{\alma}
d\al
\bigg[-{2\over\al}(m_c^2-\al(1-\al)s)
\nnb\\
&&+\int\limits_{\bemi}^{1-\al}d\be\left[(\al+\be)m_c^2-\al\be
s\right]\nnb\\
&&\times \left({1\over
\al}-{\al+\be\over2\be^2}\right)\bigg].\nnb
\enqa
\beqa
\rho_6^{\qq^2}(s)={m_c^2\rho\qq^2\over 6\pi^2}\sqrt{s-4m_c^2\over
s},
\enqa
where: $m_c,~\lag
g^2G^2\rag,~\qq, ~\mix$ are respectively the charm quark mass, gluon condensate, 
light quark and mixed condensates; 
 $\rho$ indicates the violation of the four-quark vacuum 
saturation.
The integration limits are given by:
 \bea
 \almi&=&{1\over 2}({1-v}),~~~~ \alma={1\over 2}({1+v})\nnb\\
 \bemi&=&{\al m_c^2/( s\al-m_c^2)}
 \label{eq:limit}
\eea
where $v$ is the $c$-quark velocity:
\beq
v\equiv  \sqrt{1-4m_c^2/s}~.
\eeq
\subsection{ \bf \boldmath$\lambda-J/\psi$-like molecule current} 
\nin
For the current in Eq. (\ref{eq:curr5}), the corresponding spectral functions read:
\bea\label{eq:pertlam}
\rho_\lambda^{pert}(s)&=&{1\over 2^{7}3 \pi^6}\int\limits_{\almi}^{\alma}
{d\al\over\alpha^3}
\int\limits_{\bemi}^{1-\al}{d\be\over\be^3}(1-\al-\be)\nnb\\
&&\times\left((\al+\be)m_c^2-\al\be s\right)^3\nnb\\
&&\times \bigg[(1+\al+\be)\left((\al+\be)m_c^2-\al\be s\right)
\nn\\
&&-4m_c^2(1-\al-\be)\bigg]~,
\nn\\
\rho_\lambda^{\qq}(s)&=& {\cal O}(m_q)~,
\nn\\
\rho_\lambda^{\la
G^2\ra}(s)&=&-{\la g^2G^2\ra\over2^{6}3\pi^6}\int\limits_{\almi}^{\alma} d\al\!\!
\Bigg\{\int\limits_{\bemi}^{1-\al}{d\be}\Bigg[{m_c^2(1-\al-\be)\over3\al^3}\nnb\\
&&\times\bigg[m_c^2(1-\al-\be)-\kab\nn\\
&&\times
\left(4+\al+\be+{3\over\be}(1-\al)\right)\bigg]\nnb\\
&&-{\kab\over16\al\be}\nnb\\
&&\times \left((2+\al+\be)m_c^2-\al\be s\right)
\nn\\
&&-{(1-\al-\be)\over96\al^2\be^2}\nnb\\
&&\times\kab(3-\al-\be)\Bigg]\nnb\\
&&+{(m_c^2-\al(1-\al)s)^2
\over16\al(1-\al)}\Bigg\}~,
\nnb\\
\rho_\lambda^{mix}(s)&=&{\cal O}(m_q)~,
\nn\\
\rho_\lambda^{\qq^2}(s)&=&{2\over 27\pi^2}\rho\qq^2(s+2m_c^2)\sqrt{1-{4m_c^2\over
s}}~,
\eea
where the integration limits have been defined in Eq. (\ref{eq:limit}). 
\section{ \boldmath Calibration of the method from $M_\psi$ and choice of $m_c$ } 
\nin
Using the QSSR method, one usually estimates the $J/\psi$ mass, from the ratio:
\beq
{\cal R}_{\psi}={\int_{4 m_c^2}^{t_c}ds ~s~\rho_\psi(s)~e^{-s\tau}\over\int_{4
m_c^2}^{t_c}
ds ~\rho_\psi(s)~e^{-s\tau}}\simeq M^2_{\psi}\;,
\lb{Rpsi}
\enq
where $\rho_\psi$ is the spectral density associated to the vector current:
\beq
J^\mu_\psi= \bar c\gamma^\mu c~.
\eeq
The QCD expression of the vector correlator is known in the literature \cite{svz}
including the $d=8$ condensates \cite{SNB}. The full expression of the exponential 
moments ${\cal R}_{\psi}$
is given in \cite{BELL} and its expansion in $1/m_c$ can be found in \cite{SNGh}. 
For the numerical analysis we shall  introduce the renormalization group invariant  
quantities 
$\hat\mu_q$ \cite{FNR}:
\bea
{\la\bar qq\ra}(\tau)&=&-{\hat \mu_q^3 \ga-\log{ \sqrt{\tau}\Lambda}\dr^{2/{-
\beta_1}}}\nnb\\
{\la\bar qg\sigma.Gq\ra}(\tau)&=&-{\hat \mu_q^3 \ga-\log{ \sqrt{\tau}\Lambda}
\dr^{1/{-3\beta_1}}}M_0^2~,
\eea
where $\beta_1=-(1/2)(11-2n/3)$ is the first coefficient of the $\beta$ function 
for $n$ flavours. We have used the quark mass and condensate anomalous dimensions 
reported in \cite{SNB}.  We shall use the QCD parameters in 
Table \ref{tab:param}. At the scale where we shall work, and using the parameters 
in Table \ref{tab:param}, we deduce:
\beq
\rho=2.1\pm 0.2~,
\eeq
which controls the deviation from the factorization of the four-quark condensates. 
We shall not include the $1/q^2$ term discussed in \cite{CNZ,ZAK},which is 
consistent with the LO approximation 
used here as the latter has been motivated by a phenomenological parametrization  
of the larger order terms of the QCD series.

\begin{figure}[hbt] 
\begin{center}
\centerline{\epsfig{figure=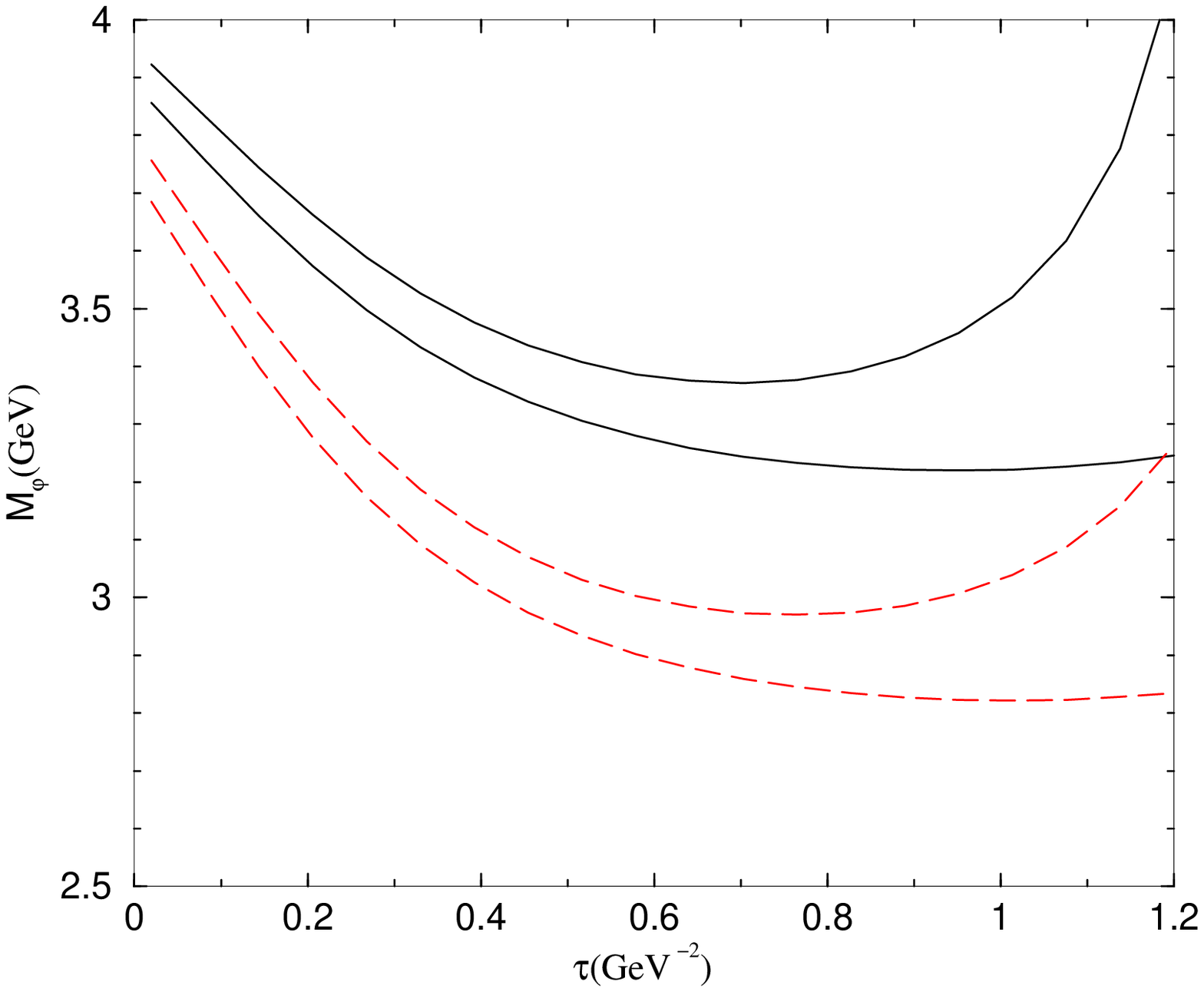,height=50mm}}
\centerline{\epsfig{figure=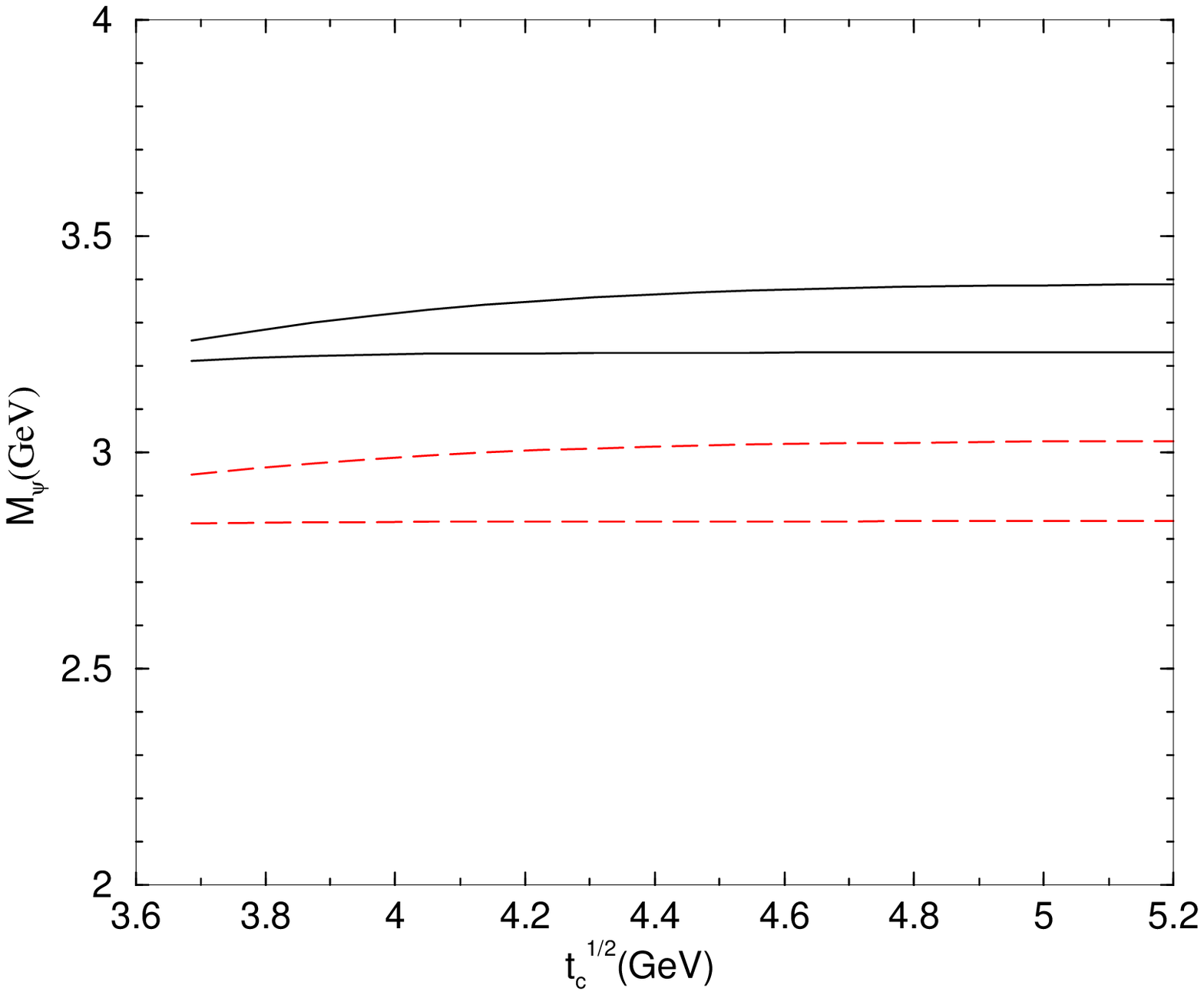,height=50mm}}
\caption{\scriptsize The $J/\psi$ mass, $M_\psi=\sqrt{{\cal R}_{\psi}}$, as a 
function
of a) $\tau$ for $\sqrt{t_c}=4.6~\GeV$ for two values of $m_c$. Solid line 
$m_c=1.47$ GeV: upper line: LO  + $\la G^2\ra$, lower line: LO +NLO +  $\la G^2\ra$. 
Dashed line: 
the same as the solid line but for $m_c=1.26$ GeV;
b) $t_c$ behaviour of $M_\psi=\sqrt{{\cal R}_{\psi}}$, 
for two different values of $m_c$. Solid line for $m_c=1.47$ GeV : 
$\tau=0.4~\GeV^{-2}$ (upper line),
and $\tau=0.8~\GeV^{-2}$ (lower line). Dashed line  for $m_c=1.26$ GeV:  
$\tau=0.4~\GeV^{-2}$ (upper line) and $\tau=0.8~\GeV^{-2}$ (lower line). }
\label{fig3} 
\end{center}
\end{figure}

{\scriptsize
\begin{table}[hbt]
\setlength{\tabcolsep}{0.2pc}
 \caption{\scriptsize    QCD input parameters. For the heavy quark masses, we use 
the range spanned
 by the running $\overline{MS}$ mass $\overline{m}_Q(M_Q)$  and the on-shell mass 
from QSSR compiled in page
 602,603 of the book in \cite{SNB}. The values of $\Lambda$ and $\hat\mu_d$ have 
been obtained from $\alpha_s(M_\tau)=0.325(8)$ \cite{SNTAU} and from the running 
masses:  $(\overline{m}_u+\overline{m}_d)(2)=7.9(3)$ MeV \cite{SNmass}. The 
original errors have been multiplied by 2 for a conservative estimate of the errors. 
    }
    {\small
\begin{tabular}{lll}
&\\
\hline
Parameters&Values& Ref.    \\
\hline
$\Lambda(n_f=4)$& $(324\pm 15)$ MeV &\cite{SNTAU,PDG}\\
$\hat \mu_d$&$(263\pm 7)$ MeV&\cite{SNmass,SNB}\\
$M_0^2$&$(0.8 \pm 0.2)$ GeV$^2$&\cite{JAMI2,HEID,SNhl}\\
$\la\alpha_s G^2\ra$& $(6\pm 2)\times 10^{-2}$ GeV$^4$&
\cite{SNTAU,LNT,SNI,fesr,YNDU,SNHeavy,BELL,SNH10,SNG}\\
$\rho\alpha_s\la \bar dd\ra^2$& $(4.5\pm 0.3)\times 10^{-4}$ GeV$^6$&
\cite{SNTAU,LNT,JAMI2}\\
$m_c$&$(1.26\sim1.47)$ GeV &\cite{SNB,SNmass,SNHmass,PDG,SNH10,IOFFE}\\
\hline
\end{tabular}
}
\label{tab:param}
\end{table}
}
\nin 
Including the $d=4$
gluon condensate, we show in Fig. \ref{fig3}a  the $\tau$-behaviour of 
$M_\psi=\sqrt{{\cal R}_{\psi}}$, for a 
given $\sqrt{t_c}=4.6$ GeV, from which the pQCD expression of the spectral density 
starts to be seen experimentally. 
\nin
From Fig.~\ref{fig3}a we see that the gluon contribution plays an important
role in stabilizing the result.
In Fig.~\ref{fig3}b  we show the $t_c$ behaviour of  $M_\psi$ for two
values of $\tau$. We see that the results are very stable against  $t_c$. 
One can deduce from  Fig.~\ref{fig3}a and Fig.~\ref{fig3}b that one can better 
reproduce the experimental 
value of  $M_{J/\psi}$ using the running mass $\overline{m}_c(m_c)$ rather than 
the on-shell mass $ m_c^{os}$. 
This feature had already been noticed in \cite{IOFFE,SNH10},  where a better 
convergence of the QCD perturbative series 
was found when working with the $\overline{MS}$ mass. Therefore, in the following 
we shall only consider the running 
mass. We have checked that the use of the on-shell mass does not affect our result 
from the double ratio of sum rules 
as it was intuitively expected.

\section{  \boldmath $M_X$ from the double ratios of sum rules (DRSR)} 
\nin
\subsection{The \boldmath$\bar 3-3$ tetraquark}
\nin
Using QSSR, one can usually estimate the mass of the $X$-meson, from the ratio $
{\cal R}_{i}$ analogue to the one in Eq. (\ref{Rpsi}),
where $i=3,6$ is related to the spectral densities obtained from the 
currents (\ref{field3}) and (\ref{field}) respectively. The $\bar 3-3$ component
of the $X$ mass has been studied  with the help of the current  (\ref{field3}). At 
the sum rule
stability point and using a slightly different (though consistent) set of QCD 
parameters than in 
Table \ref{tab:param},  one obtains with a good accuracy \cite{mnnr}:
\beq
M_3\simeq \sqrt{{\cal R}_{3}}= (3925\pm 127)~{\rm MeV}~,
\lb{m2}
\enq
and the correlated continuum threshold value fixed simultaneously by the Laplace 
and finite energy sum rules 
(FESR) sum rules:
\beq
\sqrt{t_c\vert_3}\simeq (4.15\pm 0.03)~{\rm GeV}~.
\lb{tc3}
\enq
$M_3$ is in good agreement (within the errors) with the experimental candidate 
\cite{PDG}:
\beq
M_X\vert_{exp}\simeq \sqrt{{\cal R}_{3}}= (3872.2\pm 0.8)~{\rm MeV}~,
\lb{Xexp}
\enq
while the relative low value of $t_c$ indicates that the next radial excitation 
of the $X$-meson can be in the range:
\beq
M_{X'}\approx M_X+(225\pm 127)~{\rm MeV}~.
\eeq
This low value of $t_c$ suggests that the $\bar 3-3$ resonance may be difficult to 
separate from the QCD 
continuum and  suggests also that it can be a wide resonance. 
Although  the agreement with the experimental data is remarkable, the result may 
not be sufficient
to provide a definite statement on the quark substructure of the $X$-meson. 
\subsection*{ \bd$\bar 6-6$ over $\bar 3-3$ tetraquark}
\begin{figure}[hbt] 
\begin{center}
\centerline{\epsfig{figure=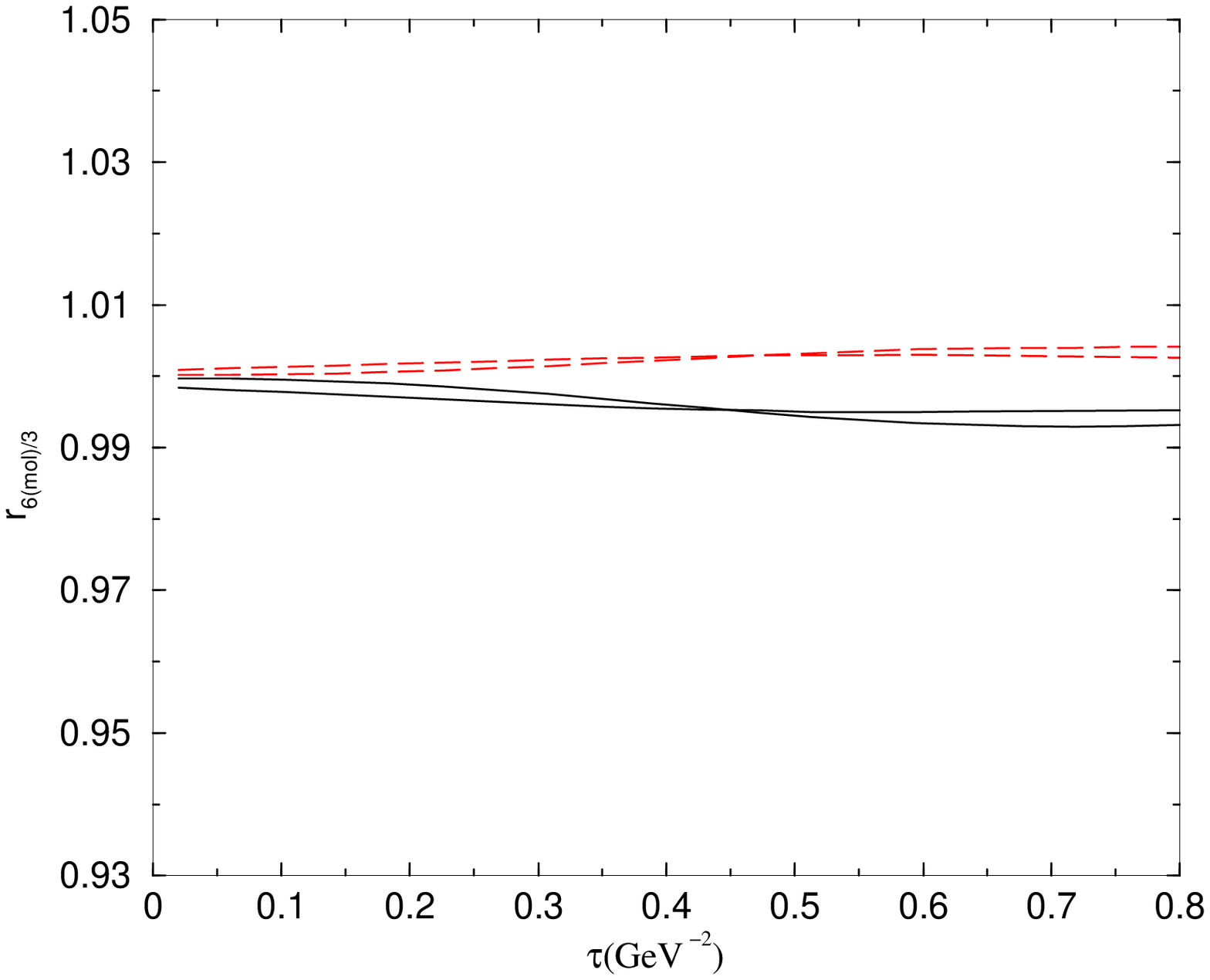,height=50mm}}
\centerline{\epsfig{figure=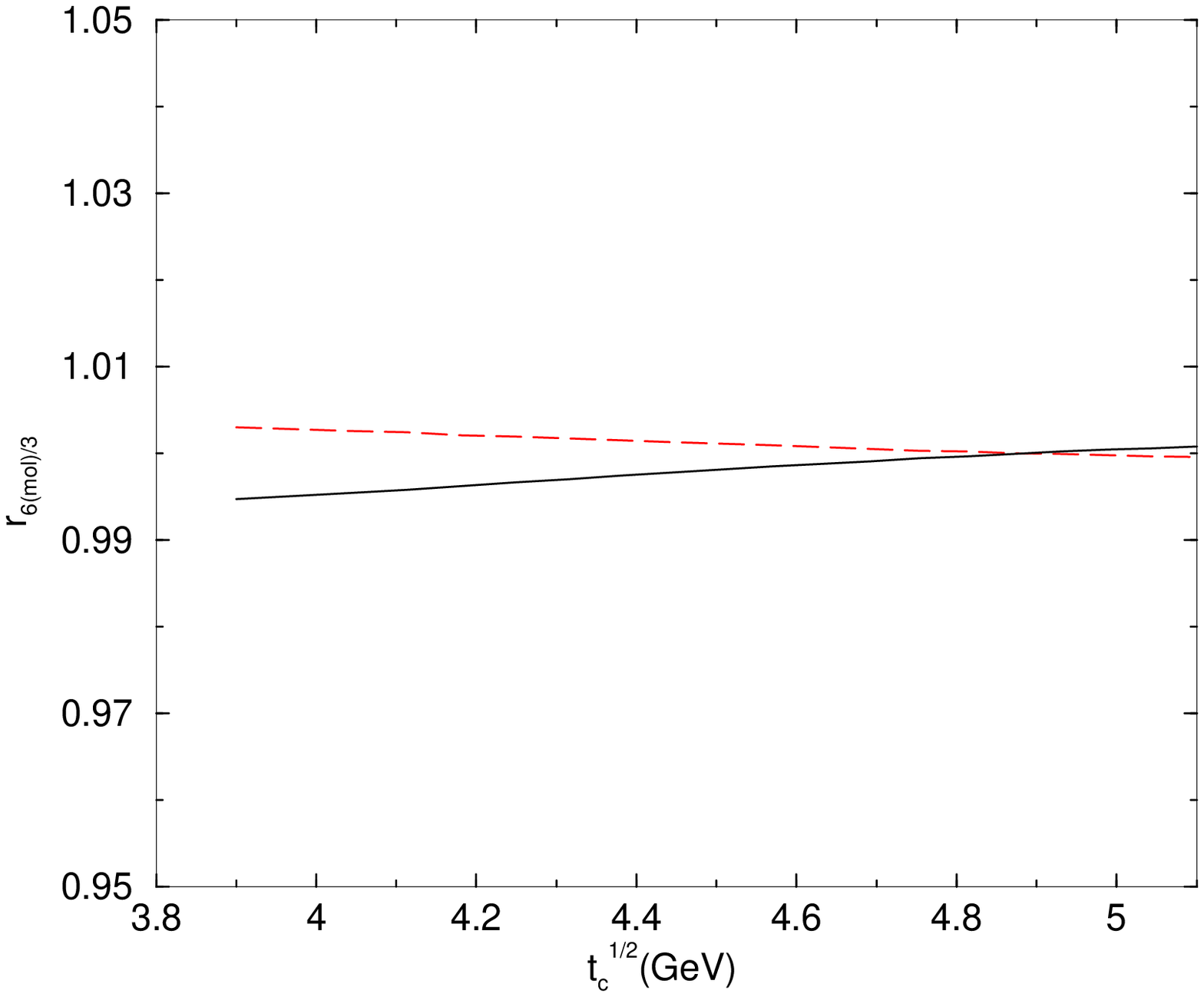,height=50mm}}
\caption{\scriptsize The double ratio $r_{6/3}$ (solid line) defined in 
Eq. (\ref{dr3}) and 
$r_{mol/3}$ (dashed line) defined in Eq. (\ref{drmol}): a)  as a function of 
$\tau$ for $\sqrt{t_c}
=4.15~\GeV$ and for two values of $m_c=1.26$ and 1.47 GeV; b) as a 
function of $t_c$ for $\tau$=0.4~GeV$^{-2}$ and $m_c=1.26$ GeV.}
\label{fig1} 
\end{center}
\end{figure}

\nin
A better understanding of the nature of the $X$, for discriminating different
proposals,  requires a more precise determination of $M_X$. This can be reached by 
considering 
the double ratio (DR) of the sum rules (DRSR)
\cite{SNGh,SNhl,SNFBS,SNFORM,SNme+e-,HBARYON}:
\beq
r_{6/3}=\sqrt{{\cal R}_{6}\over{\cal R}_{3}}\simeq {M_6\over M_3}~.
\lb{dr3}
\enq
These 
quantities are less sensitive to the choice of the heavy quark 
masses, to the perturbative radiative corrections and to the value of the 
continuum threshold than the simple ratios ${\cal R}_{\psi}$ and ${\cal R}_{3}$ 
in Eq. (\ref{Rpsi}) and~(\ref{m2}). 
Fixing $\sqrt{t_c}=4.15~\GeV$ \cite{mnnr} we show in Fig.~\ref{fig1}a the 
$\tau$-behaviour of  $r_{6/3}$ 
(continuous line) for two values of $m_c$.
One can notice that 
the result is very stable against the $\tau$-variation in a large range for 
$\tau\leq 0.8$ GeV$^{-2}$ . We show in Fig. \ref{fig1}b its $t_c$-behaviour
(continuous line) for a given $\tau=0.4$ GeV$^{-2}$ and $m_c=1.26$ GeV. We deduce:
\beq
 r_{6/3}\simeq 1.00~,
 \lb{eq:63}
 \eeq
with a negligible error, which shows that, from a QCD spectral sum rules approach, 
the $X(3872)$ can be equally described by the currents in Eqs.~ (\ref{field3}) 
and (\ref{field}).

\subsection{\bd $D^*-D$ molecule over $\bar 3-3$ tetraquark}
\nin
We can also work with the double ratio:
\beq
r_{mol/3}=\sqrt{{\cal R}_{mol}\over{\cal R}_{3}}~ .
\lb{drmol}
\enq
by using the spectral densities for the current (\ref{curr4}).  In Fig.~\ref{fig1}a 
we also show the double ratio 
$r_{mol/3}$ (dashed line) for $\sqrt{t_c}=4.15 ~\GeV$ and for two values of $m_c$, 
while we show in Fig. \ref{fig1}b  its $t_c$-behaviour
(dashed line) for a given $\tau=0.4$ GeV$^{-2}$ and $m_c=1.26$ GeV. 
One can deduce from the previous analysis:
\beq 
r_{mol/3}\simeq 1.00~,
 \lb{eq:3mol}
\eeq
also with  a negligible error.

\subsection{\bd$\lambda-J/\psi$-like-molecule over the $\bar 3-3$  tetraquark}
\nin
Using approaches similar to the previous ones, we study the ratio of the 
$\lambda-J/\psi$-like molecule over the
tetraquark $\bar 3-3$ one. We show the analysis in Fig. \ref{fig:lam-molecule} 
from which one can deduce
at the $\tau$ and $t_c$ stability regions:
\begin{figure}[hbt] 
\begin{center}
\epsfig{figure=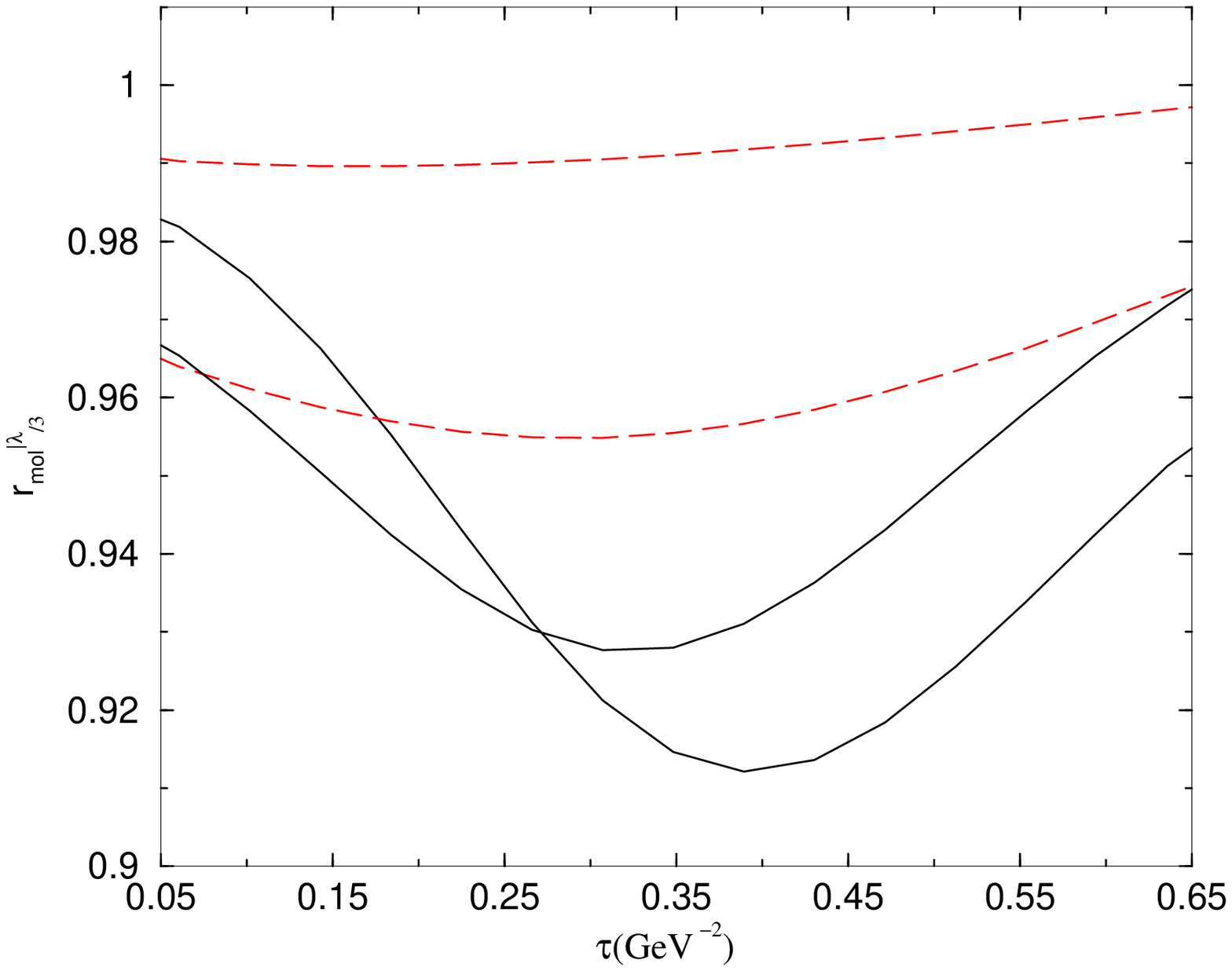,height=55mm}
\epsfig{figure=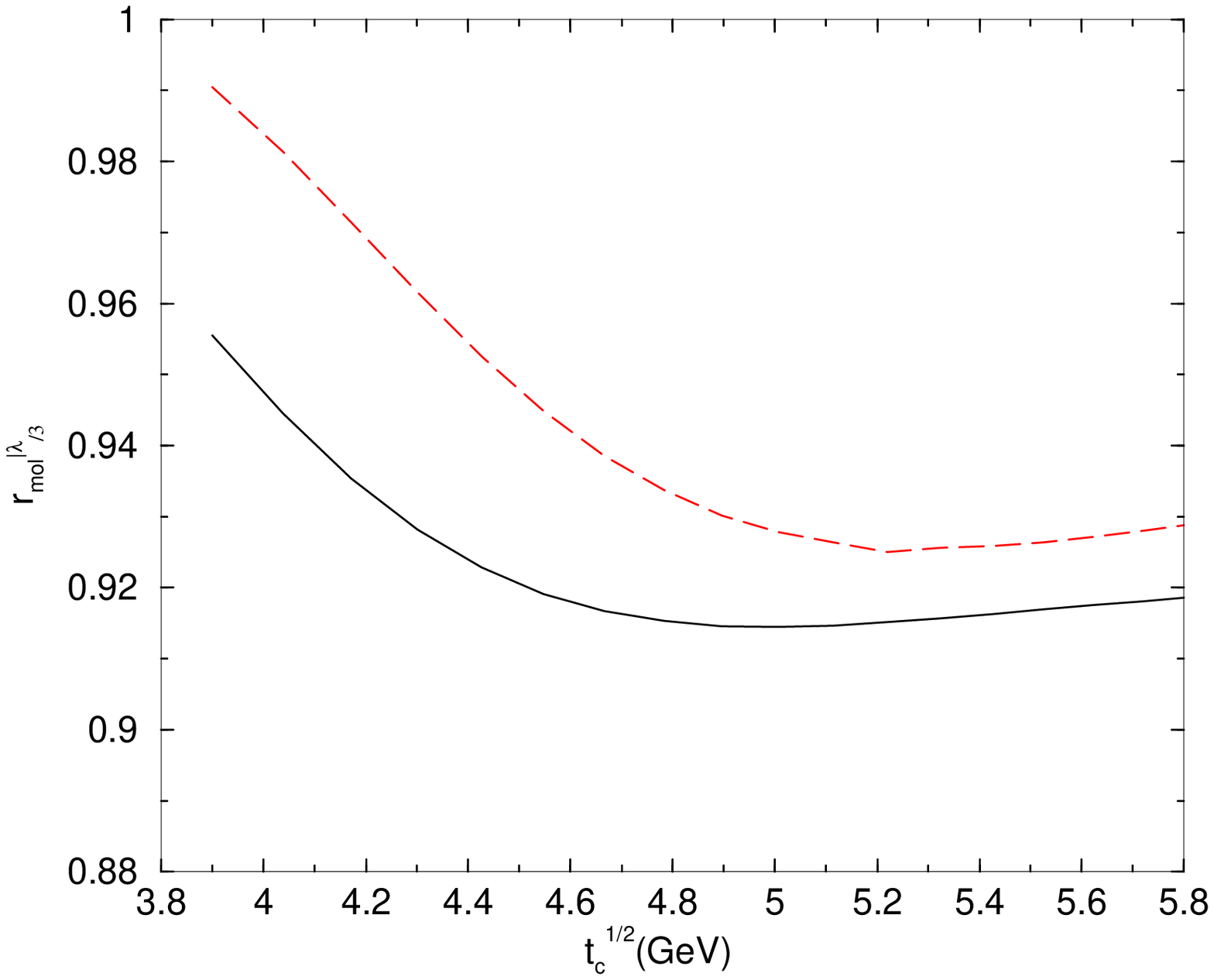,height=55mm}
\caption{The double ratios $r_{\lambda/3}$ of the $\lambda-J/\psi$-like molecule 
over the $\bar 3-3$ tetraquark masses 
defined in  (\ref{eq:ratiolam}):  a) as a function of $\tau$ for $\sqrt{t_c}= 
3.9$ GeV 
(dashed line) and 5 GeV (solid line). The upper and lower minima correspond 
respectively to 
$m_c=1.47$ and 1.26 GeV; b) as a function of $\sqrt{t_c}$ for  
$\tau$=0.35~GeV$^{-2}$ and $m_c=1.26$ 
GeV (solid line) and for $\tau$=0.3~GeV$^{-2}$ and $m_c=1.47$ GeV (dashed line) .} 
\label{fig:lam-molecule} 
\end{center}
\end{figure} 
\beq
r_{\lambda/3}\equiv {M_\lambda\over M_3}=0.96\pm 0.03~,
\label{eq:ratiolam}
\eeq
where the errors come from the stability regions and $m_c$~\footnote{The analysis 
of the ratio between 
the $\lambda$-molecule $J/\psi$-like current and $J/\psi$ mass is not conclusive 
within our approximation  
due to the absence of a stability region. The  appearance of an inflexion point  
favors a lower value 
of the $\lambda$-molecule mass. However, analyzing the ratio of the 4-quark over 
the 2-quark correlators 
which do not necessarily optimize at the same $\tau$-values may be inappropriate.}.

\subsection{Comments on the results}
\nin
\b Our analysis has shown that the three substructure assignements for the 
$X$-meson ($\bar 3-3$ and $\bar 6-6$ tetraquarks and $D-D^{(*)}$ molecule) lead
to (almost) the same mass predictions within the accuracy of the approach. 
Therefore, a priori, the alone study of the $X$-mass cannot reveal
its  nature if it is mainly composed by these substructures. \\
From the previous analysis  we observe that the distance between the
continuum threshold (about 4 GeV) and the resonance masses (see e.g. the ratio 
$r_{6/3}$ in Fig. 2)
is relatively small. This  indicates that the separation between the resonance 
and the continuum may be difficult 
to achieve. This feature is also signaled by the (almost) absence of the 
so-called sum rule window 
(a compromise region where the resonance dominates over the continuum 
contribution and where the QCD OPE is 
convergent) when one extracts the absolute mass of the $\bar 6-6$ mass. 
Then, as  in the analysis of the wide  
$\sigma$ \cite{VENEZIA} and hybrid or some other large width states 
\cite{SNB,SNH}, we expect that the $\bar 6-6$
\footnote{In a particular two-body potential model, one might expect that the   
$\bar 6-6$ tetraquark state can be 
weakly bound due to the repulsive force between the two quarks, but this may 
not  necessarily be true for a more
more general potential \cite{RICHARD}.} 
and, to a lesser extent, the $\bar 3-3$ four-quark or molecule $D-D^{(*)}$ 
states can be wide or/and weakly bound. \\
\b The analysis of the $\lambda- J/\psi$-like molecule mass in Eq. 
(\ref{eq:ratiolam}) shows that it can be lower 
than the other configurations studied previously. \\
\b In order to get a deeper understanding of the properties of these  states, 
we shall, in what follows,  compute their hadronic widths. 
\section{Can  the \bd$X$-meson hadronic width reveal its nature ?}
\nin
\begin{figure}[hbt] 
\begin{center}
\epsfig{figure=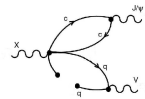,height=20mm}
\caption{Vertex diagrams contributing to the $X$-width for the diquark currents 
(\ref{field3}) and (\ref{field})  and the molecular current  (\ref{curr4}).}  
\label{fig4} 
\end{center}
\end{figure} 
One can study the decays $X\to J/\psi +3\pi$ and $X\to J/\psi+ 2\pi$  
using vertex sum rules \cite{NN}, 
where the $2\pi$ and $3\pi$ can be assumed to come from the $\rho$ and $\omega$ mesons
using vector meson dominance (VDM)\, \footnote{This approach assumes implicitly that
the decay occurs through a direct coupling of the $X$-meson to $J/\psi$ and 
$\rho,~\omega$ mesons where some eventual rescattering contributions (which 
could be important) have been neglected.}. 
In so doing, one works with the three-point function:
\bea
\Pi^{\mu\nu\alpha}(p,p',q)&\equiv& \int d^4x~d^4y ~e^{i(p'x+qy)}\times\nnb\\
&&\la 0\vert{\cal T}J_\psi^\mu(x)J_V^\nu(y)J_X^{\alpha\dagger}(0)\vert 0\ra~,
\eea
associated to the $J/\psi$-meson $J_\psi^\mu$, to the vector mesons $J_V^\nu$ 
and to the $X$-meson $J_X$.
\subsection{The tetraquarks and $D^*-D$ molecule}
 In the case of the three $X$-currents ($\bar 3-3$, $\bar 6-6$ tetraquarks and 
molecule) discussed previously, 
the lowest order and lowest dimension correction   (fall apart) QCD diagrams 
are shown in Fig. \ref{fig4}. 
An estimate of the $X-J/\psi-V$ coupling in \cite{x24,NN} indicates that if the 
$X$ is a pure $\bar 3-3$ tetraquark or a molecule state, one would obtain:
\beq
g_{X\psi\omega}^{3,mol}\simeq 14\pm 2~,
\eeq
which would correspond to a width:
\beq
\Gamma_{X\to J/\psi +n\pi}^{3,mol}\approx 50~{\rm MeV}~.
\lb{eq:gam3m}
\eeq
Doing an analogous analysis if the $X$ is a $\bar 6-6$ tetraquark state, one also 
obtains a similar value. \\
These previous results  are too big compared with the data upper bound \cite{BELLE}:
\beq
 \Gamma_{X\to {\rm all}} \leq  2.3~{\rm MeV}~.
 \label{eq:bound}
 \eeq
\subsection{The \bd$\lambda-J/\psi$-like molecule }
\nin
\begin{figure}[hbt] 
\begin{center}
\epsfig{figure=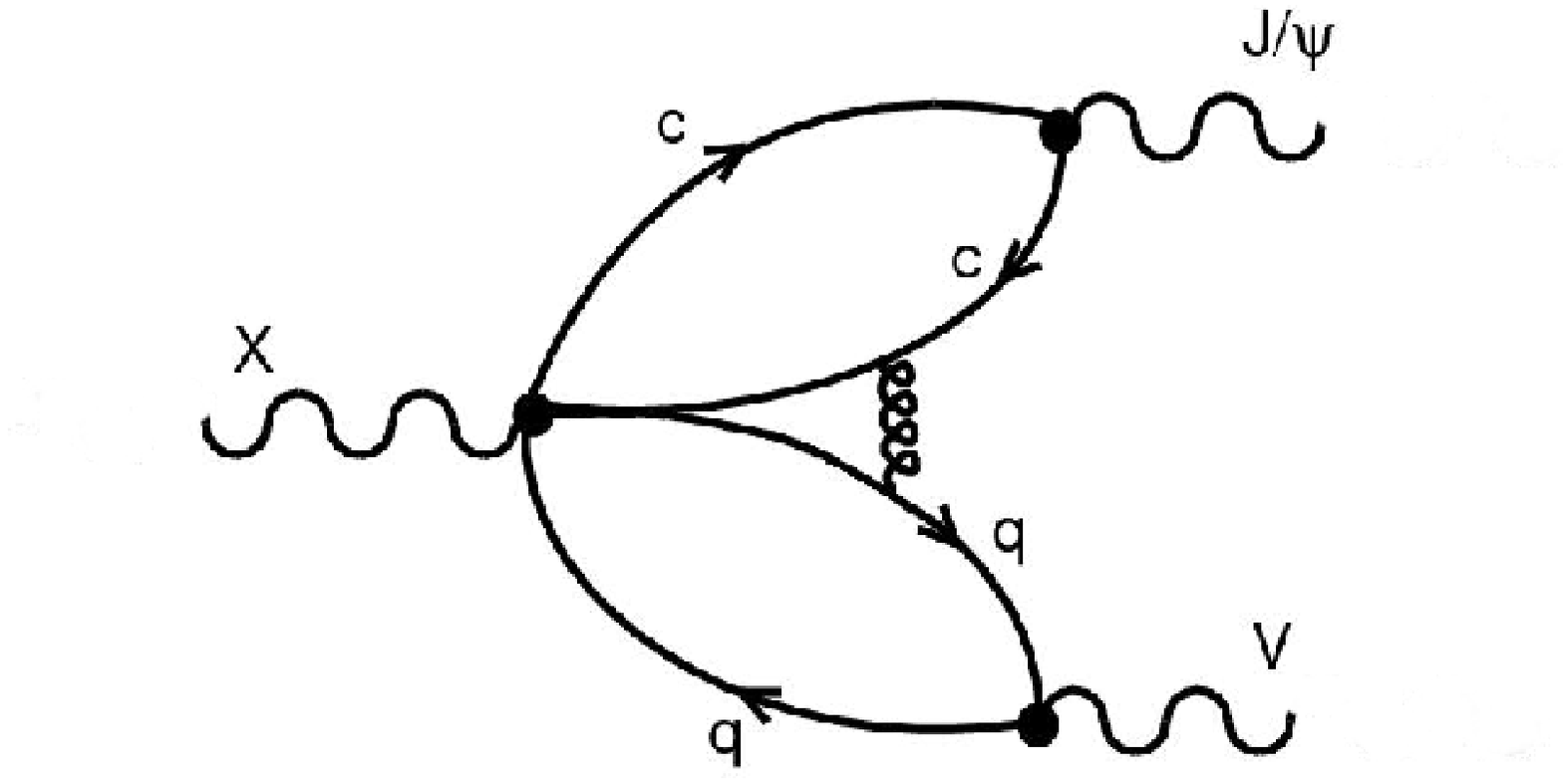,height=40mm}
\epsfig{figure=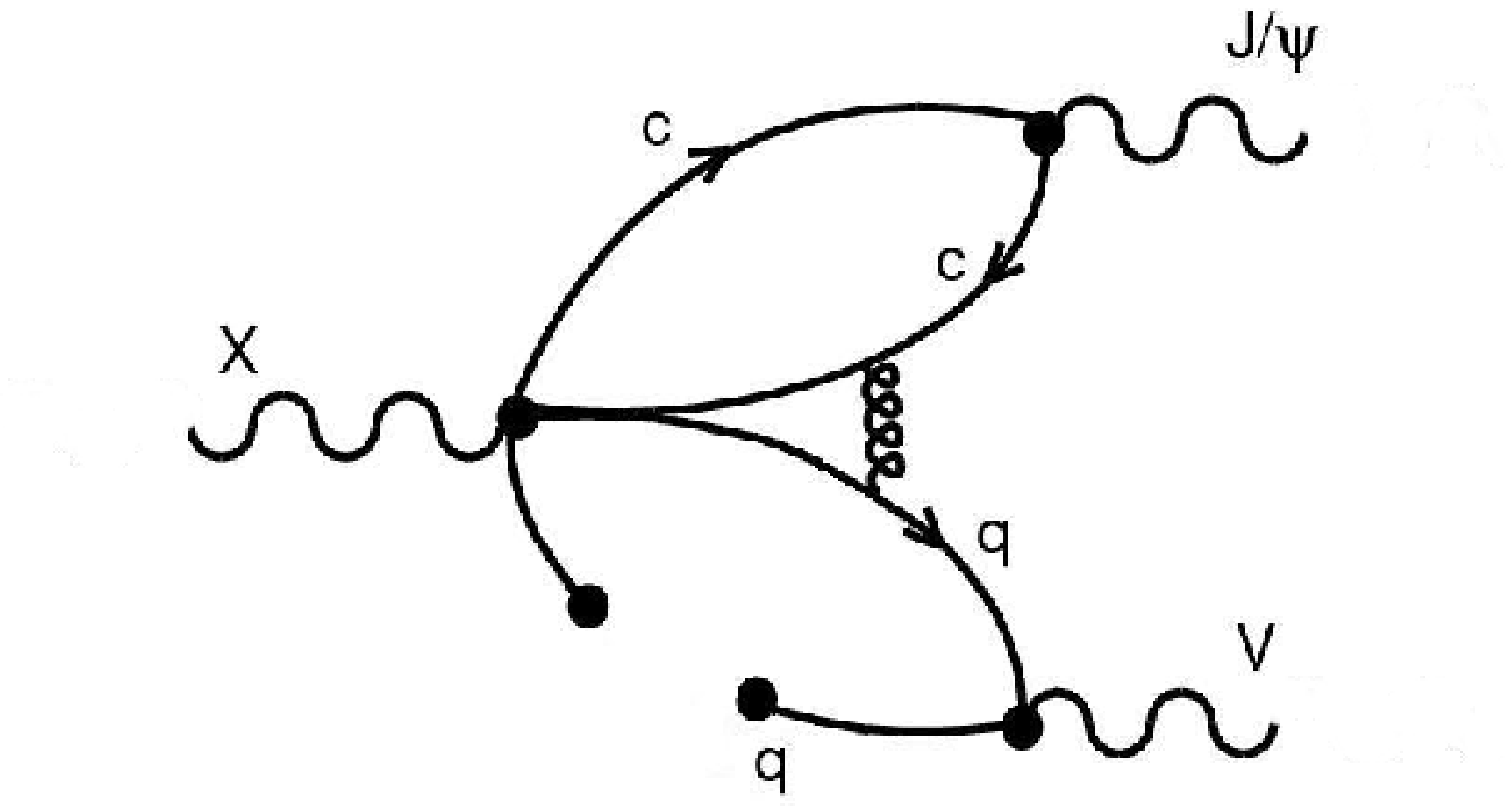,height=40mm}
\caption{Lowest order and lowest dimension Vertex diagrams contributing to the 
$X$-width for the $\lambda$-molecule current in Eq. (\ref{eq:curr5}).}  
\label{onegluon} 
\end{center}
\end{figure} 
\nin
Another possibility is to study  the $\lambda- J/\psi$-like molecule current.  
In contrast to  
to the case of previous currents, the leading order contribution to the 
three-point function
is due to one gluon exchange in Fig. \ref{onegluon}. 
The exact evaluation of these diagrams are
technically involved. However, a rough approximation by including loop 
factors\,\footnote{A similar estimate has been done in \cite{SN4} for explaining 
the too small $\gamma\gamma$ width of the $a_0(980)$ if it is a four-quark or 
molecule state.} leads to the coupling:
\beq
g^\lambda_{X\psi\omega}\approx \ga {\alpha_s\over \pi}\dr g_{X\psi\omega}^{mol}
\approx 
1~,
\eeq
where we have used $\alpha_s(M_X)\simeq 0.26$.  This would corresponds 
to a width :
\beq
 \Gamma_{X\to J/\psi +n\pi}^\lambda \approx 0.3~{\rm MeV}~,
 \label{eq:gam6}
 \eeq
which satisfies the previous experimental upper bound. Due to the rough approximation used in the estimate, we may expect that the result is known within a factor 2 .

A similar rough approximation can be made to evaluate the radiative decay width
$X(3872)\to J/\psi\gamma$. This decay was studied in ref.~\cite{ZANETTI}
considering the $X(3872)$ as having charmonium $(c\bar{c})$ and molecular
($D\bar{D}^*$) components. In the case that $X$  is a pure $\bar 3-3$ tetraquark 
or a molecule state, one would obtain within :
\beq
\Gamma_{X\to J/\psi \gamma}^{3,mol}\approx 3.4~{\rm MeV}~.
\lb{eq:gamrad}
\eeq
Therefore, using also in this case the rough approximation
\beq
\Gamma_{X\to J/\psi\gamma}^\lambda\approx \ga {\alpha_s\over \pi}\dr^2 
\Gamma_{X\to J/\psi\gamma}^{mol},
\eeq
we get within a factor 2:
\beq
 \Gamma_{X\to J/\psi\gamma}^\lambda \approx 0.02~{\rm MeV}~,
 \label{eq:gamra6}
 \eeq
which also satisfies the experimental upper bound. From the results in 
Eqs.~(\ref{eq:gam6}) and (\ref{eq:gamra6}) we would get:
\beq
 {\Gamma_{X\to J/\psi\gamma}^\lambda\over\Gamma_{X\to J/\psi\pi\pi}^\lambda}
 \approx 0.07~,
 \label{rat}
 \eeq
Taking into account the rough approximation of a factor 2 used to estimate each width, this result can be  consistent with the experimental value \cite{belle2}:
\beq
 {\Gamma_{X\to J/\psi\gamma}^{exp}\over\Gamma_{X\to J/\psi\pi\pi}^{exp}}
= 0.14\pm0.05~.
 \label{ratexp}
 \eeq
 
\section{Conclusions}
\nin
\b We have studied the mass of the $X(3872)$ using double ratios of sum rules, 
which are more accurate
than the usual simple ratios used in the literature. We found that the different 
proposed configurations  
($\bar 3-3$ and $\bar 6-6$ tetraquarks and $D-D^{(*)}$ molecule) lead to (almost) 
the same mass predictions 
within the accuracy of the method [see Eqs.  (\ref{eq:63}) and  (\ref{eq:3mol})], 
indicating that the predictions 
of the $X$-meson mass is not enough to reveal its nature. However, the (relatively) 
small distance between the 
resonance mass and the continuum threshold in the QSSR analysis and also  the 
(almost) absence of the 
sum rule window, indicate that these $\bar 3-3$ and $\bar 6-6$ tetraquarks and 
$D-D^{(*)}$ molecule states can be  
wide or weakly bound. These observations are also supported by their large 
hadronic decay widths from vertex sum rules 
analysis given in  Eq. (\ref{eq:gam3m}) \cite{NN}.  \\
\b Among these different proposals, the only eventual possibility which can lead 
to a $X(3872)$ with narrow hadronic and radiative widths consistent within the errors
with the present data,  
is the  choice of the $\lambda-J/\psi$-like molecule  current given in Eq. (\ref{eq:curr5}). \\
\b Sharper tests of the previous results can be done from an explicit evaluation of the
QCD vertex function and from a more precise experimental measurements of the
the ratio in Eq.~(\ref{ratexp}).  In this case, some eventual mixing among different currents
(see e.g. \cite{x24,ZANETTI}) may help to improve the agreement between theory and experiment.
\section{Acknowledgements}
\nin
M. Nielsen would like to thank R.D. Matheus and C.M. Zanetti for useful discussions and for some 
partial collaborations,
and the LPTA-Montpellier for the hospitality where 
this work has been initiated. This work has been partly supported 
by the CNRS-FAPESP program,  by  CNPq-Brazil and by the CNRS-IN2P3 
within the project Non-perturbative QCD and Hadron Physics.


\begin{references}
\bibitem{PDG} For a review, see e.g. PDG 08, C. Amsler et al., {\it Phys. Lett.} 
{\bf B 667} (2008) 1.
\bibitem{BELLE}  S.-K. Choi  {\it et al.} [Belle Collaboration],   
Phys. Rev. Lett. {\bf 91}, 262001 (2003).
\bibitem{BABAR}  B.~Aubert {\it et al.}  [{\sc {\sc BaBar}} Collaboration],
  Phys.\ Rev.\ D {\bf 71}, 071103 (2005).
\bibitem{CDF}D. Acosta {\it et al.} [CDF II Collaboration], Phys. Rev. Lett. 
{\bf 93}, 072001 (2004);
\bibitem{D0} 
 V.~M.~Abazov {\it et al.}  [D0 Collaboration],
  Phys.\ Rev.\ Lett.\  {\bf 93}, 162002 (2004);

\bibitem{Swanson}
 For reviews, see e.g.,  E.~S.~Swanson,
  Phys.\ Rept.\  {\bf 429}, 243 (2006); J.~M.~Richard, talk given at QCD 05 
(Montpellier 4-8th July 2005), hep-ph/0601043.
\bibitem{rev} For a review, see e.g. M. Nielsen, F. S. Navarra, S. H. Lee, 
arXiv:0911.1958.
\bibitem{svz} M.A. Shifman, A.I. and Vainshtein and V.I. Zakharov,
Nucl. Phys. {\bf B147}, 385 (1979).

\bibitem{rry} L.J. Reinders, H. Rubinstein and S. Yazaki, Phys. Rept. 
{\bf 127}, 1 (1985). 

\bibitem{SNB} For a review and references to original works, see
e.g., S. 
Narison, {\it QCD as a theory of hadrons,
Cambridge Monogr. Part. Phys. Nucl. Phys. Cosmol.} {\bf 17}, 1-778
(2002) 
[hep-h/0205006]; {\it QCD
spectral sum rules ,  World Sci. Lect. Notes Phys.} {\bf 26}, 1-527
(1989);
{ Acta Phys. Pol.} {\bf B26} (1995) 687; { Riv. Nuov. Cim.} {\bf 10N2} 
(1987) 1; { Phys. Rept.} {\bf 84}, 263 (1982).

\bibitem{mnnr} R.D. Matheus, S. Narison, M. Nielsen, J.M. Richard,
Phys. Rev. {\bf D75}, 014005 (2007).

\bibitem{RICHARD} See e.g. K. Terasaki, arXiv:1005573 [hep-ph];
J.M. Richard, QCD 10 (July 2010-Montpellier) and private comunication.

\bibitem{SN4} S. Narison, Phys. Lett. {\bf B175}, 88 (1986).

\bibitem{x24} R. D. Matheus, F. S. Navarra, M. Nielsen, C. M. Zanetti,
Phys. Rev. {\bf D80}, 056002 (2009).

\bibitem{BELL}J.S. Bell and R.A. Bertlmann, {\it Nucl. Phys.} {\bf B 227}, (1983) 
435; R.A. Bertlmann, {\it Acta Phys. Austriaca} {\bf 53}, (1981) 305;  R.A. Bertlmann 
and H. Neufeld, {\it Z. Phys.} {\bf C 27} (1985) 437.
\bibitem{SNGh} S. Narison, {\it Phys. Lett.} {\bf B 387} (1996) 162.
 \bibitem{FNR}E.G. Floratos, S. Narison and E. de Rafael, {\it Nucl. Phys.} 
{\bf B 155} (1979) 155.
 \bibitem{CNZ} K. Chetyrkin, S. Narison and V.I. Zakharov, {\it Nucl. Phys.} 
{\bf B 550} (1999)
353;  S. Narison and V.I. Zakharov, {\it Phys. Lett.} {\bf B 522},  (2001) 266; 
S. Narison and V.I. Zakharov, {\it Phys. Lett.} {\bf B 679},  (2009) 355.
\bibitem{ZAK} For reviews, see e.g.: V.I. Zakharov, {\it Nucl. Phys. Proc. Suppl.} 
{\bf 164} (2007) 240; S. Narison,  {\it Nucl. Phys. Proc. Suppl.} {\bf 164},  225 (2007). 





\bibitem{SNTAU}S. Narison, {\it Phys. Lett.} {\bf B 673} (2009) 30.

 \bibitem{SNmass} For reviews, see e.g.: S. Narison, {\it Phys.Rev.} {\bf D 74} 
(2006) 034013; 
 arXiv:hep-ph/0202200;  {\it Phys. Lett.} {\bf B 216} (1989) 191; {\it Phys. Lett.} 
{\bf B 358} (1995) 113; {\it Phys. Lett.} {\bf B 466} (1999) 345; S. Narison, H.G. 
Dosch, {\it Phys. Lett.} {\bf B 417} (1998) 173; 
S. Narison, N. Paver, E. de Rafael and D. Treleani, {\it Nucl. Phys.} {\bf B 212} 
(1983) 365; S. Narison, E. de Rafael, {\it Phys. Lett.} {\bf B 103} (1981) 57; C. 
Becchi, S. Narison, E. de Rafael, F.J. Yndurain, {\it Z. Phys.} {\bf C 8} (1981) 335.

\bibitem{JAMI2}Y. Chung et al.{\it Z. Phys.} {\bf C 25} (1984) 151;  H.G. Dosch, 
Non-Perturbative Methods (Montpellier 1985);  
H.G. Dosch, M. Jamin and S. Narison, {\it Phys. Lett.} {\bf B  220} (1989) 251.
\bibitem{HEID}B.L. Ioffe, {\it Nucl. Phys.} {\bf B 188} (1981) 317 , {\bf
B 191} (1981) 591; A.A.Ovchinnikov and A.A.Pivovarov,
{\it Yad.\ Fiz.}  {\bf 48} (1988) 1135.
\bibitem{SNhl}S. Narison, Phys. Lett. {\bf B 210}, 238 (1988); Phys. Lett. 
{\bf B 605}, 319 (2005).
\bibitem{LNT}G. Launer, S. Narison and R. Tarrach, {\it Z. Phys.} {\bf C 26} (1984) 433.
\bibitem{SNI}S. Narison, {\it Phys. Lett.} {\bf B 300} (1993) 293; ibid {\bf B 361} 
(1995) 121.
\bibitem{fesr} R.A. Bertlmann, G. Launer and E. de Rafael, 
{\it Nucl. Phys.} {\bf B 250} (1985) 61; R.A. Bertlmann et al., 
{\it Z.\ Phys.}  {\bf C 39} (1988) 231.
\bibitem{YNDU}F.J. Yndurain, hep-ph/9903457.
\bibitem{SNHeavy}S. Narison, {\it Phys. Lett.} {\bf B 387} (1996) 162.
\bibitem{SNH10}S. Narison,  arXiv:1004.5333 [hep-ph] and references therein.
\bibitem{SNG} S. Narison, { Phys. Lett.} {\bf B361}, 121 (1995);
 {Phys.Lett.} {\bf B624} (2005) 223.
\bibitem{SNHmass}S. Narison, {\it Phys. Lett.} {\bf B 197} (1987) 405; 
{\bf Phys. Lett.} {\bf B 341} (1994) 73; {\it Phys. Lett.} {\bf B 520} (2001) 115.



\bibitem{IOFFE} B.L. Ioffe and K.N. Zyablyuk, {\it Eur. Phys. J.} {\bf  C 27} (2003) 
229; B.L. Ioffe, {\it Prog. Part. Nucl. Phys.} {\bf 56} (2006) 232.




\bibitem{SNFBS} S. Narison, {\it Phys. Lett.} {\bf B 322} (1994) 327.
\bibitem{SNFORM} S. Narison, {\it Phys. Lett.} {\bf B 337} (1994) 166; {\it Phys. Lett.} 
{\bf B 668} (2008) 308.
\bibitem{SNme+e-} S. Narison, {\it Phys.Rev.} {\bf D 74} (2006) 034013; 
{\it Phys. Lett.} {\bf B 358} (1995) 113; {\it Phys. Lett.} {\bf B 466} (1999) 34.
\bibitem{HBARYON}R.M. Albuquerque, S. Narison, M. Nielsen, {\it Phys. Lett.} 
{\bf B 684} (2010) 236;
R.M. Albuquerque, S. Narison, {\it Phys. Lett.} {\bf B 694} (2010) 217.
\bibitem{VENEZIA}S. Narison, G. Veneziano, {\it Int. J. Mod. Phys.}
 {\bf A 4} (1989) 2751; S. Narison, {\it Nucl. Phys.} {\bf B 509} (1998) 312; 
{\it Phys. Rev.} {\bf D 73} (2006) 114024.  

\bibitem{SNH}S. Narison, {\it Phys. Lett.} {\bf B 675} (2009) 319.


\bibitem{NN}F. Navarra and M. Nielsen, {\it Phys. Lett.} {\bf B 639} (2006) 272.

\bibitem{ZANETTI}M. Nielsen, C.M. Zanetti, arXiv:1006.0467 [hep-ph].


 
\bibitem{belle2} 
  K.~Abe {\it et al.} [Belle Collaboration], hep-ex/0505037, 
hep-ex/0505038. 
\end{references}
\end{document}